\documentclass[11pt]{article}

\usepackage{amsmath}
\usepackage{amssymb}
\usepackage{graphicx}
\usepackage[numbers,compress]{natbib}
\begin{document}
	
	\title{Operators of Dirac's theory with mass and  axial chemical potential}	

\author{Ion I. Cot\u aescu\thanks{Corresponding author E-mail:~~i.cotaescu@e-uvt.ro},\\
	{\it West University of Timi\c soara,} \\{\it V. Parvan Ave. 4,
		RO-300223 Timi\c soara}}

\maketitle

\begin{abstract}
	The Dirac equation with mass and axial chemical potential is solved analytically obtaining the mode spinors and  corresponding projection operators giving the spectral representations of the principal conserved operators. In this framework, the odd partner of the Pryce spin operator is defined for the first time showing how these operators may be combined for defining the particle and antiparticle spin and polarization operators of Dirac's theory of massive fermions either in the free case or in the presence of the axial chemical potential.   The quantization procedure is applied in both these cases obtaining two distinct operator algebras in which the particle and antiparticle spin and polarization operators take canonical forms. In this approach  statistical operators with independent particle and antiparticle vortical chemical potentials may be constructed.

\end{abstract}


\section{Introduction}

For understanding the rotation effects in quark-gluon plasma one releases on the   helicity operator of Dirac's theory \cite{B,V1,V2,V3} or even on the projection of the spin operator on a given direction (see for instance \cite{Buz} and the literature indicated therein). The helicity operator is odd having negative charge conjugation parity such that   after quantization its particle and antiparticle terms have opposite signs. On the other hand, we have shown recently \cite{Cot} that the natural conserved spin operator of this theory is that proposed by  Pryce according to his hypothesis (e) \cite{Pr}. This is an even operator which allowed us to define an even operator of fermion polarization, which  is different from the helicity one \cite{Cot,Cot1}.  We consider this discrepancy as an opportunity because this could be the starting point  to an approach in which the particle and antiparticle polarization operators may be defined independently.   The principal goal of this paper is to write down the conserved operators of Dirac's theory, including those related to the spin and polarization effects, either in the free case or  in the presence of chemical potentials. As in the relativistic quantum mechanics (RQM) one cannot interpret the antiparticle terms,  we finally construct the corresponding one-particle operators of the quantum field theory (QFT) we need for writing down suitable  statistical operators. 

In the case of the Dirac free fields we derived already the even polarization operator that can be combined with the helicity one for obtaining the particle and antiparticle polarization operators \cite{Cot}. As this operator is related to the Pryce(e) spin operator,  we may ask which is the odd spin-type operator corresponding to the helicity one. Solving this problem  with the help of the spectral representations we find a new odd vector-operator whose components do not satisfy a $su(2)$ algebra but combined with the components of the Pryce(e) operator  form a larger $o(4)$ algebra. This  can be split in two independent $su(2)$ algebras generated by the components of the particle and antiparticle spin operators we define here for the first time. 

For extending these results to the theory with chemical potentials  we start with the Dirac equation of massive fermions  having only the vector and axial chemical potentials which may be seen as a new equation as long as the axial term does not commute with the mass term of the Hamiltonian operator.  Solving this equation we obtain new mode spinors depending on the axial chemical potential and the corresponding projection operators we need for constructing the spectral representations of  the conserved operators. We observe that in the presence of the axial chemical potential the Pryce(e) spin  and its associated odd spin-type operator are conserved in spite of the fact that  the axial term breaks  the spin symmetry defined in Ref. \cite{Cot}. Combining these operators we  define  the particle and antiparticle spin and polarization operators giving their spectral representations.  We obtain thus a coherent and flexible approach in which the vortical effects can be studied separately for particles and antiparticles.  The principal benefits of this approach are the mode spinors of Dirac's theory with mass  and axial chemical potential as well as the form of the projection operators which depend only on this chemical potential.  Based on these results, the conserved one-particle operators of QFT  are derived shoving how the field operators of the theory of free field are related to those derived in Dirac's theory with mass and axial potential. 

We start in the next section revisiting the mode spinors of the momentum basis of the Dirac free field, focusing on the projection operators of this basis. The next section is devoted to the spin and polarization operators. Here we define the odd spin-type operator shoving that its components and those of the Pryce(e) spin operator form a $so(4)$ algebra that can be split  in two independent $su(2)$ algebras of particle and antiparticle spin operators. Moreover, the particle and antiparticle polarization operators are defined combining our polarization operator derived in Ref. \cite{Cot} with the helicity one. This section ends with ta brief inspection of the momentum-helicity basis and its projection and polarization operators. Section 4 is devoted to the Dirac equation with axial chemical potential which is analytically solved obtaining new mode spinors and corresponding projection operators we use for constructing the new spin and polarization operators in the presence of the axial chemical potential. In the next section we perform the quantization obtaining the operator algebras of the systems without or with axial chemical potential. We show that these systems are different, their field operators being related through a non-trivial Bogolyubov transformation. In the last section we comment on how the statistical operators may be constructed in the theory with mass and axial chemical potential.

In three Appendices we present successively, the algebra of Dirac's matrices and the $SL(2,\mathbb{C})$ generators, the polarization spinors and the coefficients of the Bogolyiubov transformations considered here.   

\section{Momentum basis and its projection operators}

The Dirac field $\psi   :M\times {\cal V}_P\to {\cal V}_D$  depend on coordinates of the Minkowski space-time $M$ and on the arbitrary Pauli spinors $\xi\in {\cal V}_P$ defining the fermion polarization. For this reason we  use two equivalent  notations for the Dirac free field,  $\psi \equiv \psi _{\xi}$, and similarly for the mode spinors,  the second notation being useful when we study the role of the polarization spinors \cite{Cot,Cot1}. 

The Dirac free fields of mass $m$ are solutions of the  Dirac equation $E_D\psi   =(i\gamma^{\mu}\partial_{\mu}-m)\psi   =0$ that can be put in Hamiltonian form,   
\begin{equation}\label{hamD}
	i\partial_t\psi   (x)=H_D\psi   (x)\,, \quad 	H_D=-i\gamma^0\gamma^i\partial_i+m\gamma^0\,.	
\end{equation}  	
These solutions form the space ${\cal F}=\{\psi   \,|\, E_D\psi   =0\}$ equipped with  the Dirac relativistic scalar product 
\begin{equation}\label{sp}
	\langle\psi   ,\psi   '\rangle_D=\int d^3x\overline{\psi   }(x)\gamma^0\psi   '(x)=\int d^3x \psi   ^ {+} (x)\psi   '(x)	\,.
\end{equation}
In momentum representation the free fields  $\psi   \in {\cal F}$ may be expanded  as \cite{BDR,KH}
\begin{eqnarray}\label{Psi}
	\psi     (x)&=&	\psi     ^+(x)+	\psi     ^-(x)\nonumber\\
	&=&\int d^3p \sum_{\sigma}\left[U_{{\bf p},\sigma}(x) \alpha_{\sigma}({\bf p}) +V_{{\bf p},\sigma}(x) \beta^ {*} _{ \sigma}({\bf p})\right]\,,~~~~
\end{eqnarray}
in terms of  mode spinors,  $U_{{\bf p},\sigma}$ and  $V_{{\bf p},\sigma}=C  U_{{\bf p},\sigma}^*$, of positive and respectively negative frequencies,   related through the charge conjugation defined by the matrix $C=C^{-1}=i\gamma^2$.  The 
spinors-functions $\alpha: \Omega_{\mathring p}\to {\cal V}_P$ and   $\beta: \Omega_{\mathring p}\to {\cal V}_P$ are the  particle and respectively antiparticle {\em wave spinors} defined  on the orbit $ \Omega_{\mathring p}$ of mass $m$ having the representative momentum   $\mathring p=(m,0,0,0)$. The expansion (\ref{Psi}) shows that the set ${\frak B}=\{U_{{\bf p},\sigma}, V_{{\bf p},\sigma}\}$ is a generalized basis of the space ${\cal F}$  that can be split  in two  subspaces, of positive and respectively negative frequencies, ${\cal F}={\cal F}^+\oplus {\cal F}^-$, which are orthogonal with respect to the scalar product (\ref{sp}). Note that after quantization the wave spinors become particle or antiparticle field operators. 

Solving  the Dirac equation one obtains  the mode spinors 
\begin{eqnarray}
	U_{{\bf p},\sigma}(x)&=&u({\bf p},\sigma)\frac{ e^{-iE_pt+i{\bf x}\cdot{\bf p}}}{(2\pi)^{\frac{3}{2}}}\,,\label{U1}\\
	V_{{\bf p},\sigma}(x)&=&v({\bf p},\sigma)\frac{ e^{iE_pt-i{\bf x}\cdot{\bf p}}}{(2\pi)^{\frac{3}{2}}}\,,\label{V1}
\end{eqnarray}
where  $E_p=\sqrt{m^2+p^2}$ (with $p=|{\bf p}|$) is the usual relativistic energy. In the standard representation of the $\gamma$-matrices, the spinors $u$ and $v=Cu^*$  have the form
\begin{eqnarray}
	u({\bf p},\sigma)&=&\sqrt{\frac{m}{E_p}}l_{\bf p}\left(  
	\begin{array}{c}
		\xi_{\sigma}({\bf n})\\
		0
	\end{array}\right)\,,
	\nonumber\\  
	&=&	\frac{1}{\sqrt{2E_p(E_p+m)}}\left(  
	\begin{array}{c}
		(E_p+m)\,\xi_{\sigma}({\bf n})\\	
		p^i\sigma_i  \,\xi_{\sigma}({\bf n})
	\end{array}\right)\,,\label{u}\\
	v({\bf p},\sigma)&=& \sqrt{\frac{m}{E_p}}l_{\bf p} \left(  
	\begin{array}{c}
		0\\
		-\eta_{\sigma}({\bf n})	
	\end{array}\right)\,,
	\nonumber\\
	&=&	\frac{1}{\sqrt{2E_p(E_p+m)}}\left(  
	\begin{array}{c}
		-p^i\sigma_i \,\eta_{\sigma}({\bf n})\\
		-(E_p+m)\,\eta_{\sigma}({\bf n})	
	\end{array}\right)\,,\label{v}
\end{eqnarray}
determined by the Wigner boost  (\ref{Ap}) and the  related polarization spinors, $\xi_{\sigma}({\bf n})$  and $\eta_{\sigma}({\bf n})=i\sigma_2 \xi_{\sigma}({\bf n})^*$ defined in the Appendix B. These spinors give the fermion polarization $\sigma=\pm\frac{1}{2}$ through the eigenvalue problems 
\begin{eqnarray}
	\hat{\bf s}\cdot{\bf n}\,\xi_{\sigma}({\bf n})	=\sigma\, \xi_{\sigma}({\bf n}) \,, \quad 
	\hat{\bf s}\cdot{\bf n}\,\eta_{\sigma}({\bf n})	=-\sigma\, \eta_{\sigma}({\bf n})\,,\label{xy}
\end{eqnarray}
of the  projection of  the Pauli spin operator (\ref{si}) along the unit vector ${\bf n}$.  In general,  ${\bf n}$ may depend on ${\bf p}$ when we say that the polarization is {\em peculiar}. Otherwise, when ${\bf n}$ is independent on ${\bf p}$, we speak about a {\em common} polarization.  

The spinors $u$ and $v$ have the properties
\begin{eqnarray}
	u^+({\bf p},\sigma)	u({\bf p},\sigma')&=&	v^+({\bf p},\sigma)	v({\bf p},\sigma')	=\delta_{\sigma\sigma'}\,,\label{ortu}\\
	u^+({\bf p},\sigma)	v(-{\bf p},\sigma')&=&	v^+(-{\bf p},\sigma)	u({\bf p},\sigma')=0\,,\label{ortv}	
\end{eqnarray}
leading to the orthonormalization 
\begin{eqnarray}
	\langle U_{{\bf p},\sigma}, U_{{{\bf p}\,}',\sigma'}\rangle_D &=&
	\langle V_{{\bf p},\sigma}, V_{{{\bf p}\,}',\sigma'}\rangle_D=	\delta_{\sigma\sigma^{\prime}}\delta^{3}({\bf p}-{\bf p}\,^{\prime})\,,\label{ortU}\\
	\langle U_{{\bf p},\sigma}, V_{{{\bf p}\,}',\sigma'}\rangle_D &=&
	\langle V_{{\bf p},\sigma}, U_{{{\bf p}\,}',\sigma'}\rangle_D =0\,, \label{ortV}
\end{eqnarray}
and completeness,
\begin{eqnarray}
	&&	\int d^3p \sum_{\sigma}\left[ U_{{\bf p},\sigma}(t,{\bf x})U_{{\bf p},\sigma}^{+}(t,{\bf x}\,')+V_{{\bf p},\sigma}(t,{\bf x})V_{{\bf p},\sigma}^{+}(t,{\bf x}\,')\right] \nonumber\\
	&&\hspace*{20mm}=\delta^3({\bf x}-{\bf x}\,')\,,\label{comp}
\end{eqnarray}
of the basis of mode spinors. Hereby we deduce the inversion formulas
\begin{equation}\label{inv}
	\alpha_{\sigma}({\vec p})=\langle U_{{\vec p},\sigma},\psi \rangle_D\,, \quad 
	\beta_{\sigma}({\vec p})=\langle \psi,  V_{{\vec p},\sigma} \rangle_D\,, 
\end{equation}
we need in applications. 

The principal operators acting on ${\cal F}$ are multiplicative, differential or Fourier operators. The last ones are integral operators  acting as 
\begin{equation}\label{Y1}
	(A \psi   )(t,{\bf x})=\int d^3x' {\frak A}(t,{\bf x}-{\bf x}^{\,\prime})\psi   (t,{\bf x}^{\,\prime})\,,
\end{equation}
through local  kernels preserving  the time  and allowing  three-dimensional Fourier representations, 
\begin{equation}\label{KerY0}
	{\frak A}(t,{\bf x}) =\int d^3p\,\frac{e^{i {\bf p}\cdot{\bf x}}}{(2\pi)^3}  {\hat A}(t,{\bf p})\,,
\end{equation} 
depending on the matrices  $\hat A(t,{\bf p})\in{\rho}_D$  we call  the Fourier transforms of the operators $A$.  Then the action (\ref{Y1}) on a field  (\ref{Psi})  can be written as
\begin{eqnarray}
	(A \psi   )(t, {\bf x})&=&	\int d^3x'\, {\frak A}(t,{\bf x}-{\bf x}^{\,\prime})\psi   (t,{\bf x}^{\,\prime})\nonumber\\
	&=&\int d^3p \sum_{\sigma}\left[\hat A(t,{\bf  p})U_{{\bf p},{\sigma}}(t, {\bf x})\alpha_{\sigma}({\bf p})\right.\nonumber\\
	&&\left.~~~~~~+\hat A(t,-{\bf  p})V_{{\bf p},{\sigma}}(t, {\bf x})\beta^*_{\sigma}({\bf p})\right]\,.\label{Y2}
\end{eqnarray}
One can verify that a Fourier operator $A$ is self-adjoint  with respect to the scalar product (\ref{sp}) if its Fourier transform is a Hermitian matrix, $\hat A(t,{\bf  p})=\hat A(t,{\bf  p})^+$. 

There are many  Fourier operators whose kernels are independent on time.  Moreover, the differential operator involving only space derivatives also can be seen as Fourier operators. An  example is the Dirac Hamiltonian  (\ref{hamD}) which has the Fourier transform 
\begin{equation}\label{HDp}
	\hat H_D({\bf p})=m\gamma^0+\gamma^0 \boldsymbol\gamma \cdot {\bf p} \,.
\end{equation}
Other elementary examples are the momentum-independent matrices of $\rho_D$, $\gamma^{\mu}$, $s_{\mu\nu}$,...etc., defined in the Appendix A,  that can be seen as Fourier operators whose Fourier transforms are just the matrices themselves. 

Having the form of the mode spinors we may define the Hermitian projection operators $\Pi_{\sigma}^{(\pm)}$ on the subspaces of particles $(+)$ and antiparticle $(-)$ of polarizations $\sigma$ giving their Fourier transforms 
\begin{eqnarray}
	\Pi_{\sigma}^{(+)}~~~\Rightarrow~~~	\hat\Pi^{(+)}_{\sigma}({\bf p})&=&u({\bf p},\sigma)u({\bf p},\sigma)^+ \,, \\
	\Pi_{\sigma}^{(-)}~~~\Rightarrow~~~	\hat\Pi^{(-)}_{\sigma}({\bf p})&=&v(-{\bf p},\sigma)v(-{\bf p},\sigma)^+ \,,
\end{eqnarray}
which are related through charge conjugation as
\begin{equation}
	\hat\Pi^{(-)}_{\sigma}({\bf p})=C \left[\hat\Pi^{(+)}_{\sigma}(-{\bf p})\right]^*C\,.	
\end{equation}
The action of these operators 
\begin{eqnarray}
	( \Pi^{(+)}_{\sigma} U_{{\bf p},\sigma'})(x)&=&\hat\Pi^{(+)}_{\sigma}({\bf p})U_{{\bf p},\sigma'}(x)=\delta_{\sigma\sigma'}U_{{\bf p},\sigma}(x)\,,\nonumber\\
	( \Pi^{(-)}_{\sigma} U_{{\bf p},\sigma'})(x)&=&\hat\Pi^{(-)}_{\sigma}({\bf p})U_{{\bf p},\sigma'}(x)=0\,,\nonumber	\\	
	( \Pi^{(+)}_{\sigma} V_{{\bf p},\sigma'})(x)&=&\hat\Pi^{(+)}_\sigma(-{\bf p})V_{{\bf p},\sigma'}(x)=0\,,\nonumber\\
	( \Pi^{(-)}_{\sigma} V_{{\bf p},\sigma'})(x)&=&\hat\Pi^{(-)}_{\sigma}(-{\bf p})V_{{\bf p},\sigma'}(x)=\delta_{\sigma\sigma'}V_{{\bf p},\sigma}(x)\,,\nonumber
\end{eqnarray}  
shows that thy form a complete system of orthogonal projection operators satisfying
\begin{eqnarray}
	\Pi^{(+)}_{\sigma} \Pi^{(+)}_{\sigma'}=\delta_{\sigma\sigma'} \Pi^{(+)}_{\sigma}\,,&\quad& 	\Pi^{(-)}_{\sigma} \Pi^{(-)}_{\sigma'}=\delta_{\sigma\sigma'} \Pi^{(-)}_{\sigma}\,,
	\nonumber\\\
	\Pi^{(\pm)}_{\sigma} \Pi^{(\mp)}_{\sigma'}=0\,,&\quad& \sum_{\sigma}\left(\Pi^{(+)}_{\sigma}+\Pi^{(-)}_{\sigma}\right)=I\,.
\end{eqnarray}
These projection operators have complicated expressions depending on the form of the polarization spinors. With their help we may write the Pryce projection operators  $\Pi ^{(\pm)}=\sum_{\sigma}\Pi^{(\pm)}_{\sigma}$ on the subspaces  of positive and negative frequencies,  $\Pi^{(+)}{\cal F}={\cal F}^+$ and  $\Pi^{(-)}{\cal F}={\cal F}^-$ \cite{B}, whose Fourier transforms, 
\begin{eqnarray}
	\hat\Pi^{(+)}({\bf p})&=&\sum_{\sigma}\hat\Pi^{(+)}_{\sigma}({\bf p})\nonumber\\
	&=&\frac{m}{E_p}l_{\bf p}	\frac{1+\gamma^0}{2} l_{\bf p}=\frac{1}{2}\left(1+\frac{\hat H_D({\bf p})}{E_p}\right)\,,\label{Pip}\\ 
	\hat\Pi^{(-)}({\bf p})&=&\sum_{\sigma}\hat\Pi^{(-)}_{\sigma}({\bf p}) \nonumber\\
	&=&\frac{m}{E_p}l^{-1}_{\bf p}	\frac{1-\gamma^0}{2} l^{-1}_{\bf p}=\frac{1}{2}\left(1-\frac{\hat H_D({\bf p})}{E_p}\right)\,,~~~~\label{Pim}
\end{eqnarray}
are independent on the form of the polarization spinors. Hereby we see that the operator  (\ref{HDp}), can be represented now as 
\begin{equation}\label{HDp1}
	\hat H_D({\bf p})=E_p\left[\hat \Pi^{(+)}({\bf p})-\hat\Pi^{(-)}({\bf p})\right]=E_p \hat N({\bf p})	\,,
\end{equation}
defining the Fourier transform, $\hat N({\bf p})$, of the new operator $N=\Pi^{(+)}-\Pi^{(-)}$ 
which becomes the operator of number of particles after quantization.

\section{Even and odd spin and polarization  operators}

The basis of polarization spinors  can be changed at any time, $\xi\to \hat r\xi$, by  applying a rotation $\hat r \in SU(2)$ which change the form of the Dirac spinor  giving rise to the new representation ${T}^s:\hat r\to {T}^s_{\hat r}$ of the group $SU(2)$ which encapsulates the  spin symmetry.  The operators of this representation have the action     
\begin{equation}\label{Rs}
	\left(	{T}^s_{\hat r(\theta)}\psi_{\xi}\right)(x)=\psi _{\hat r(\theta)\xi}(x)\,,
\end{equation} 
where $\hat r(\theta)$ are the rotations (\ref{r}) with Cayley-Klein parameters. The components of the conserved spin operator of Dirac's theory can be defined now as the generators of this representation, \cite{Cot}
\begin{equation}\label{Spipi}
	S_i=\left.i\frac{\partial T^s_{\hat r(\theta)}}{\partial \theta^i}\right|_{\theta^i=0} 	~~\Rightarrow~~ S_i\psi _{\xi}=\
	\psi_{\hat s_i\xi}\,.
\end{equation}
We have shown that these are Fourier operators whose Fourier transforms read \cite{Cot},
\begin{eqnarray}
	{\hat S}_i({\bf p})&=&\frac{1}{2}\sum_{\sigma,\sigma'}\left[ u({\bf p},\sigma)	\Sigma_{i\,\sigma,\sigma'}({\bf n})u^+({\bf p},\sigma')  \right. \nonumber\\
	&&\left. \hspace*{14mm} -v(-{\bf p},\sigma)	\Sigma^*_{i,\sigma,\sigma'}({\bf n})v^+(-{\bf p},\sigma')\right]\nonumber\\
	&=&{s}_i({\bf p}) \, \hat\Pi_+({\bf p}) +  {s}_i ( -{\bf p})\,\hat \Pi_-({\bf p}) \label{SSS}	\,,	 
\end{eqnarray} 
where 	$\Sigma_{i\,\sigma,\sigma'}({\bf n})=\xi^+_{\sigma}({\bf n})\sigma_i \xi_{\sigma}({\bf n})$ are the matrix elements of the matrices $\Sigma_i({\bf n})$ given by Eqs. (\ref{Sigp1}) of the Appendix B. The components 
\begin{equation}\label{sCh}
	{s}_i({\bf p})=l_{\bf p} {s}_i\,  l_{\bf p}^{-1} =\frac{E}{m}s_i-\frac{p^i\, {\bf p}\cdot {\bf s}}{m(E+m)}-\frac{i}{2m}\epsilon_{ijk}\,p^j\gamma^0\gamma^k  \,,
\end{equation}
are of the spin operator proposed by  Chakrabarti \cite{Ch}  but which is not conserved. This operator has the properties
\begin{equation}\label{sCh1}
	{\bf s}({\bf p})={\bf s}\,^+(-{\bf p})\,, \quad {\bf s}(\pm{\bf p})\hat\Pi_{\pm}({\bf p})=\hat\Pi_{\pm}({\bf p}) {\bf s}(\mp{\bf p})\,,
\end{equation}
allowing us to obtain the final expressions 
\begin{eqnarray}
	{\hat S}_i({\bf p})=\frac{m}{E} {s}_i+\frac{{p}^i\, {\bf s}\cdot{\bf p}}{E(E+m)}+\frac{i}{2E}\epsilon_{ijk}{p}^j {\gamma}^k \,,\label{PrS}	
\end{eqnarray}
which are  just the operators proposed by Pryce according to his hypothesis (e) \cite{Pr}.  

The form of the spin operator allowed us to define the operator of fermion polarization  for any  related polarization spinors, $\xi_{\sigma}({\bf n})$  and $\eta_{\sigma}({\bf n})$, satisfying the general eigenvalues problems (\ref{xy}).
The corresponding polarization operator may be defined as the Fourier operator $W_s$ whose Fourier transform reads \cite{Cot}
\begin{eqnarray}
	\hat W_s({\bf p})=w({\bf p})\hat\Pi_+({\bf p}) +  w(-{\bf p}) \hat\Pi_-({\bf p})
	\,,	\label{Pol}
\end{eqnarray}
where $w({\bf p})={\bf s}({\bf p})\cdot {\bf n}({\bf p})$. As in the case of the spin operator we find that the operator of fermion polarization acts as \cite{Cot}
\begin{eqnarray}
	(W_s U_{{\bf p},\sigma})(x)&=&\hat W_s({\bf p})U_{{\bf p},\sigma}(x)=\sigma U_{{\bf p},\sigma}(x)\,,\label{WU}\\
	(W_s V_{{\bf p},\sigma})(x)&=&\hat W_s({\bf p})V_{{\bf p},\sigma}(x)=-\sigma V_{{\bf p},\sigma}(x)\,,\label{WV}
\end{eqnarray}
allowing the spectral representation\begin{equation}
	W_s=\sum_{\sigma}\sigma\left(\Pi^{(+)}_{\sigma}-\Pi^{(-)}_{\sigma}\right)\,.
\end{equation}
This operator completes the system  of commuting operators $\{P_{\mu}=i \partial_{\mu}, W_s\}$	  defining the momentum  bases as the set of common eigenspinors satisfying the eigenvalue problems,   
\begin{eqnarray}
	& i\partial_t U_{{\bf p},\sigma}=E U_{{\bf p},\sigma}\,,\quad ~~~~
	&~~~	 i\partial_t V_{{\bf p},\sigma}=-E V_{{\bf p},\sigma}\,,\\
	&-i \partial_j U_{{\bf p},\sigma}={p}^j\, U_{{\bf p},\sigma}\,,\quad ~~~~~
	&-i \partial_j V_{{\bf p},\sigma}=-{p}^j\, V_{{\bf p},\sigma}\,,\\
	&{W}_s U_{{\bf p},\sigma}=\sigma\, U_{{\bf p},\sigma}\,,\quad ~~~
	&~~~	{W}_s V_{{\bf p},\sigma}=-\sigma\, V_{{\bf p},\sigma}\,,
\end{eqnarray}
All these operators as well as the spin components are even operators which after quantization have positive charge parity in the sense  the particle and antiparticle terms of the corresponding one-particle operator have the same signs.

However, this convenient approach is quite new because only odd polarization operators were considered so far, as the time-like component of the Pauli-Lubanski operator or the helicity operator proposed recently \cite{V1,V2}.  For understanding the relations among these operators and our even operators it deserves to study  all the odd operators related to the spin and polarization. 

We observe first that if $A$ is an even operator then we can define its odd associated  operator $A^{\flat}=AN$. Therefore, we may introduce here for the first time  the odd spin-type operator  ${\bf S}^{\flat}={\bf S}N$ whose components have the Fourier transforms 
\begin{eqnarray}
	\hat S_i^{\flat}({\bf p})&=&	\hat S_i ({\bf p})\hat N({\bf p})={s}_i({\bf p}) \, \hat\Pi_+({\bf p}) -  {s}_i ( -{\bf p})\,\hat \Pi_-({\bf p})	\nonumber\\
	&=&\gamma^0 s_i-\frac{	p^i \gamma^0 {\bf p}\cdot{\bf s}}{E(E+m)}+\frac{p^i}{2E}\gamma^5	\,.
\end{eqnarray} 
Similarly, we define the odd polarization operator $W_s^{\flat}=W_sN$ having the Fourier transform
\begin{eqnarray}
	\hat W^{\flat}_s({\bf p})&=&\hat W_s({\bf p})\hat N({\bf p})=\sum_{\sigma}\sigma\left(\Pi^{(+)}_{\sigma}+\Pi^{(-)}_{\sigma}\right)\nonumber\\
	&=&	w({\bf p})\hat\Pi_+({\bf p})-  w(-{\bf p}) \hat\Pi_-({\bf p})
	\,.	\label{Po1l} 	
\end{eqnarray}
The operators $S_i^{\flat}$ and $W_s^{\flat}$ are Hermitian, conserved and translation invariant but the odd spin-type operator is no longer a genuine spin one generating a $su(2)$ algebra. Nevertheless, its components and those of the spin operator generate a $so(4)\sim su(2)\times su(2)$ Lie  algebra satisfying the commutation rules
\begin{eqnarray}
	\left[ S_i  ,	 S_j\right]&=&i\epsilon_{ijk} S_k \,,\\
	\left[ S_i  ,	 S^{\flat}_j\right]&=&i\epsilon_{ijk} S_k^{\flat} \,,\\	
	\left[ S^{\flat}_i  ,	 S^{\flat}_j\right]&=&i\epsilon_{ijk} S_k\,. 	
\end{eqnarray} 
Hereby,  we may define new particle (pa) and antiparticle (ap) spin operators  
\begin{eqnarray}
	S_i^{pa}=\frac{1}{2}\left(S_i+S_i^{\flat}  \right)\,, \quad
	S_i^{ap}=\frac{1}{2}\left(S_i-S_i^{\flat}  \right)\,, 	
\end{eqnarray}
having the Fourier transforms
\begin{eqnarray}
	\hat S_i^{pa}({\bf p})&=&\frac{1}{2}\left(\hat S_i({\bf p})+\hat S_i^{\flat}({\bf p})  \right)=s_i({\bf p})\hat\Pi^{(+)}({\bf p})\,,\\ 
	\hat S_i^{ap}({\bf p})&=&\frac{1}{2}\left(\hat S_i({\bf p})-\hat S_i^{\flat}({\bf p})  \right)=s_i(-{\bf p})\hat\Pi^{(-)}({\bf p})\,. 	
\end{eqnarray}
These operators are Hermitian conserved and translation invariant  generating two independent $su(2)$ algebras.  Similarly we define the particle and antiparticle  polarization operators                                              
\begin{eqnarray}
	W_s^{pa}=\frac{1}{2}\left(W_s+W_s^{\flat}  \right)\,, \quad
	W_s^{ap}=\frac{1}{2}\left(W_s-W_s^{\flat}  \right)\,, 	
\end{eqnarray}
whose Fourier transforms
\begin{eqnarray}
	\hat W_s^{pa}({\bf p})&=&\frac{1}{2}\left(\hat W_s({\bf p})+\hat W_s^{\flat}({\bf p})  \right)=w({\bf p})\hat\Pi^{(+)}({\bf p})\,,\\ 
	\hat W_s^{ap}({\bf p})&=&\frac{1}{2}\left(\hat W_s({\bf p})-\hat W_s^{\flat}({\bf p})  \right)=w(-{\bf p})\hat\Pi^{(-)}({\bf p})\,, 	
\end{eqnarray}
guarantee that these are Hermitian conserved and translation invariant operators commuting between themselves. Note that in the case of common polarization, when ${\bf n}$ is independent on momentum, we recover the traditional definitions 
\begin{eqnarray}
	W_s={\bf n}\cdot{\bf S}\,, \quad W^{\flat}_s={\bf n}\cdot{\bf S}^{\flat}\,,
\end{eqnarray}
which explain why we need to introduce  the new odd spin-type operator ${\bf S}^{\flat}$.  

{\em The momentum-helicity basis},  largely used in applications, is the only example of peculiar polarization we have until now.  In this basis, in which the spin is projected on the momentum direction, ${\bf n}_p=\frac{{\bf p}}{p}$, we denote the polarization spinors simpler as $\xi({\bf p})$ and $\eta({\bf p})$ bearing in mind that now $\sigma$ is the helicity. The spinors (\ref{u}) and (\ref{v}) that read now
\begin{eqnarray}
	u_h({\bf p},\sigma)&=&\frac{1}{\sqrt{2E_{p} (E_{p} +m)}}\left(  
	\begin{array}{c}
		(E_{p} +m)\xi_{\sigma}({\bf p})\\
		2\sigma p\xi_{\sigma}({\bf p})
	\end{array}\right)\,,\label{uh}\\
	v_h({\bf p},\sigma)&=&\frac{1}{\sqrt{2E_{p} (E_{p} +m)}}\left(  
	\begin{array}{c}
		2\sigma p\eta_{\sigma}({\bf p})\\
		-(E_{p} +m)\eta_{\sigma}({\bf p})	
	\end{array}\right)\,,\label{vh}
\end{eqnarray}
may be substituted in Eqs. (\ref{U1}) and (\ref{V1}) obtaining the mode spinors $U_{h\,{\bf p},\sigma}$ and  $V_{h\,{\bf p},\sigma}$ which constitute   the momentum-helicity basis ${\frak B}_h$. When we use this basis we denote the free field as $\psi_h$.

The advantage of the basis ${\frak B}_h$ is that the coressponding projection operators take more comprehensive forms  having the Fourier transforms
\begin{eqnarray}
	\hat \Pi^{(\pm)}_{h\,\sigma}({\bf p})&=&\left( 1\pm\frac{m}{E}\gamma^0 \right)\left( \frac{1}{4}+\sigma \frac{{\bf p}\cdot{\bf s}}{p}\right)\nonumber\\
	&\pm&\frac{1}{4 E}\left( \gamma^0{\bf \gamma}\cdot{\bf p}+2\sigma\,p\,\gamma^5 \right)	\,.
\end{eqnarray}
The Fourier transforms of the polarization operators can be written now as  
\begin{eqnarray}
	W_h ~\Rightarrow~	\hat W_h({\bf p})&=&\sum_{\sigma}\sigma\left(\hat\Pi_{\sigma}^{(+)}({\bf p}) -\hat\Pi^{(-)}_{\sigma}({\bf p})  \right)\nonumber\\
	&=&\frac{m\gamma^0}{E}\frac{{\bf s}\cdot{\bf p}}{p}+\frac{p  }{2 E} \gamma^5=\frac{1}{p}\,{\bf p}\cdot {\hat{\bf S}^{\,\flat}}({\bf p})\,,\label{p1}\\
	W_h^{\flat}	~\Rightarrow~	\hat W^{\flat}_h({\bf p})&=&\sum_{\sigma}\sigma\left(\hat\Pi_{\sigma}^{(+)}({\bf p}) +\hat\Pi^{(-)}_{\sigma}({\bf p})  \right)\nonumber\\
	&=&\frac{{\bf s}\cdot{\bf p}}{p}=\frac{1}{p}\,{\bf p}\cdot {\hat{\bf S}}({\bf p})=\hat h({\bf p})\,,
\end{eqnarray}
where $\hat h({\bf p})$ is the Fourier transform of the   helicity operator $W^{\flat}_h=h$ \cite{V1,V2}. The even polarization operator (\ref{p1}), we present here for the first time, is related to the chiral one as in the massless limit we have $\lim_{m\to0}W_h=\frac{1}{2} \gamma^5$. Therefore the particle and antiparticle polarization operators of the massless fermions,
\begin{equation}
	\lim_{m\to 0}	W_h^{pa/ap}=\frac{1}{4}\,\gamma^5\pm \frac{1}{2}\,h\,, 
\end{equation}
combine the helical and chiral contributions. 

\section{Dirac's theory   with mass and axial chemical potential}

The Dirac equation of massive fermions with vector (V),  axial (A)  and helical (H) chemical potentials considered until now \cite{V2}
\begin{equation}\label{1}
	\left(	i\gamma^{\mu}\partial_{\mu}-m +\mu_V\gamma^0+\mu_A\gamma^0\gamma^5 +\mu_H \gamma^0 h \right)\psi   (x)=0\,,
\end{equation}
involves the helical term $\mu_H \gamma^0 h$ given by the odd operator $h$. As mentioned before, one of our  goals is to generalize this term adding the contribution of an even polarization operator. However, for deriving this operator as in the case of the free fields we need to know the mode spinors and the projector operators of the momentum basis. 

Therefore, we  start solving  the Dirac equation with $\mu_H=0$  that can be put in Hamiltonian form 
\begin{eqnarray}\label{HA}
	i\partial_t \psi_A  =H_A\psi_A   \,,\quad H_A= -i\gamma^0\gamma^i\partial_i+m\gamma^0-\mu_A\gamma^5\,, 
\end{eqnarray}
after performing the substitution $\psi  (x)=\psi _A (x) e^{i\mu_V t}$.  The general solutions of  this equation, $\psi_A \in {\cal F}_A\sim{\cal F}$,  may be expanded in momentum representation,  
\begin{eqnarray}\label{PsiA}
	\psi_A   (x)&=&\psi_A  ^{+}(x)+\psi_A  ^{-}(x)\nonumber\\
	&=&	\int d^3p \sum_{\sigma}\left[U_{A,{\bf p},\sigma}(x) \hat\alpha_{\sigma}({\bf p}) +V_{A,{\bf p},\sigma}(x) \hat\beta^ {*} _{ \sigma}({\bf p})\right],
\end{eqnarray} 
assuming that the mode spinors of negative frequencies are related to the positive frequency ones through the charge conjugation, $V_{A,{\bf p},\sigma}={ C}U^*_{A, {\bf p},\sigma}$,  just as in the free case. Furthermore,  after a few manipulation we find the mode spinors, 
\begin{eqnarray}
	U_{A,{\bf p},\sigma}(x)&=&u_A({\bf p},\sigma)\frac{ e^{-iE_{p\sigma} t+i{\bf x}\cdot{\bf p}}}{(2\pi)^{\frac{3}{2}}}\,,\\
	V_{A,{\bf p},\sigma}(x)&=&v_A({\bf p},\sigma)\frac{ e^{iE_{p\sigma} t-i{\bf x}\cdot{\bf p}}}{(2\pi)^{\frac{3}{2}}}\,,
\end{eqnarray}
expressed in terms of new spinors in momentum representation,
\begin{eqnarray}
	u_A({\bf p},\sigma)&=&\frac{1}{\sqrt{2E_{p\sigma} (E_{p\sigma} +m)}}\left(  
	\begin{array}{c}
		(E_{p\sigma} +m)\xi_{\sigma}({\bf p})\\
		(2\sigma p-\mu_A)\xi_{\sigma}({\bf p})
	\end{array}\right)\,,\label{coco}\\
	v_A({\bf p},\sigma)&=&\frac{1}{\sqrt{2E_{p\sigma} (E_{p\sigma} +m)}}\left(  
	\begin{array}{c}
		(2\sigma p-\mu_A)\eta_{\sigma}({\bf p})\\
		-(E_{p\sigma} +m)\eta_{\sigma}({\bf p})	
	\end{array}\right)\,.\label{moco}
\end{eqnarray}
corresponding to the energies  \cite{V2}         
\begin{equation}\label{Eps}
	E_{p\sigma} =\sqrt{m^2+(2\sigma p- \mu_A)^2}=E_p-\frac{2\sigma p}{E_p}\mu_A+{\cal O}(\mu_A^2)\,,
\end{equation}  
which depend explicitly on $\mu_A$ and  polarization. The spinors $u_A$ and $v_A$ are normalized satisfying similar relations as (\ref{ortu}) and (\ref{ortv}) and, consequently,  the mode spinors $U_A$ and $V_A$ satisfy  similar  orthonormalization and completeness relations as in the free case, i. e.  (\ref{ortU}), (\ref{ortV}) and (\ref{comp}). Therefore, the set ${\frak B}_A=\{ U_{A,{\bf p},\sigma},\,V_{A,{\bf p},\sigma}\}$ represents a new "rotated" basis in ${\cal F}={\cal F}_A^+\oplus {\cal F}_A^-$, giving a new  splitting of ${\cal F}$ in subspaces of spinors of positive and negative frequencies, ${\cal F}_A^{\pm}\not={\cal F}^{\pm}$. More specific, the  transformation between the basis ${\frak B}_A$ and that of the free case, ${\frak B}_h$, has  non-vanishing particle-antiparticle matrix elements, $\langle U_{A,{\bf p},\sigma}, V_{h\,{\bf p},\sigma}\rangle\propto\mu_A$ as we shall see later.

The above solutions can be obtained only if  the polarization spinors  are those of the helicity basis such that  there are no longer  spin degrees of freedom and  the spin symmetry is broken.  Nevertheless, conserved even and odd spin operators which  commute with the Hamiltonian (\ref{HA})  can be constructed generalizing the definition (\ref{SSS}) while  the conserved polarization operators  can be defined using spectral representations in terms of projection operators. As in the previous case, the new projection operators $\Pi^{(\pm)}_{A\,\sigma}$  may be defined giving their  Fourier transforms, 
\begin{eqnarray}
	\Pi^{(+)}_{A\,\sigma}~~~\Rightarrow~~~\hat \Pi^{(+)}_{A\,\sigma}({\bf p})&=&u_A({\bf p},\sigma)u_A({\bf p},\sigma)^+\,,\\
	\Pi^{(-)}_{A\,\sigma}~~~\Rightarrow~~~\hat \Pi^{(-)}_{A\,\sigma}({\bf p})&=&v_A(-{\bf p},\sigma)v_A(-{\bf p},\sigma)^+\,,
\end{eqnarray}
that, after a little calculation, can be expressed as 
\begin{eqnarray}
	\hat\Pi^{(\pm)}_{A\,\sigma}({\bf p}) &=&\left(1\pm\frac{m\gamma^0}{E_{p\sigma} }\right)\left(\frac{1}{4}+\sigma\frac{{\bf s}\cdot{\bf p}}{p}\right)\pm\left(1-2\sigma \frac{\mu_A}{p}\right)\frac{\gamma^0 \boldsymbol\gamma \cdot{\bf p}+2\sigma p\gamma^5 }{4 E_{p\sigma} } \,.\label{p}
\end{eqnarray}
These projection operators have complicated Fourier transforms but that can be manipulated using algebraic codes on computer. We may verify thus that these  operators form a complete system of orthogonal projection operators. Moreover, we may define the operators of number of particles (pa) and antiparticles (ap), 
\begin{equation}\label{nn}
	N_A^{pa}=\sum_{\sigma}\Pi_{A\,\sigma}^{(+)}=\Pi_A^{(+)}\,,\quad
	N_A^{ap}=-\sum_{\sigma} \Pi^{(-)}_{A\,\sigma}=- \Pi_A^{(-)}	\,,
\end{equation} 
and that of the total number of particles $N_A=N_A^{pa}+N_A^{ap}$ whose  Fourier transform
\begin{equation}
	\hat N_A({\bf p})=\sum_{\sigma}\left(\hat \Pi_{A\,\sigma}^{(+)}({\bf p}) -\hat\Pi^{(-)}_{A\,\sigma}({\bf p})\right) \,,	
\end{equation}
is no longer proportional with that of the Hamiltonian operator (\ref{HA}) which now can be represented as
\begin{eqnarray}
	\hat H_A({\bf p})&=&\gamma^0\boldsymbol\gamma\cdot {\bf p}+m\gamma^0-\mu_A \gamma^5\nonumber\\
	& =& \sum_{\sigma}E_{p\sigma} \left(\hat\Pi_{A\,\sigma}^{(+)}({\bf p}) -\hat \Pi^{(-)}_{A\,\sigma}({\bf p}) 
	\right)\,,		
\end{eqnarray}
since for $\mu_A\not=0$ the energy (\ref{Eps}) depends on polarization.

The particle and antiparticle conserved spin operators, ${\bf S}^{pa/ap}_A$, may be defined directly assuming that the  Fourier transforms of their components are 
\begin{eqnarray}
	{\hat S}^{pa}_{A\,i}({\bf p})&=&\frac{1}{2}\sum_{\sigma,\sigma'}\left[ u_A({\bf p},\sigma)	\Sigma_{i\,\sigma,\sigma'}({\bf p})u_A^+({\bf p},\sigma')  \right] \,,\\
	{\hat S}^{ap}_{A\,i}({\bf p})&=&-\frac{1}{2}\sum_{\sigma,\sigma'}\left[ v_A(-{\bf p},\sigma)	\Sigma_{i\,\sigma,\sigma'}(-{\bf p})v_A^+(-{\bf p},\sigma')  \right]  	\,,	 
\end{eqnarray} 
where  the matrix elements
\begin{equation}\label{Sig}
	\Sigma_{i\,\sigma,\sigma'}({\bf p})=\xi^+_{\sigma}({\bf p})\sigma_i\, \xi_{\sigma}({\bf p}) 	
\end{equation}
depend now on the Pauli spinors of the helicity basis.  These operators have very complicated and less relevant forms but can be manipulated exploiting only the orthogonality of the spinors $u_A$ and $v_A$. We verify thus that the total spin operator ${\bf S}_A={\bf S}^{pa}_A +{\bf S}_A^{ap}$ acts on the field $\psi_A$ just as in Eq. (\ref{Spipi}) but without generating a spin symmetry. Moreover, we find that they are conserved and translation invariant forming two independent $su(2)$ algebras. 

The new even and odd conserved polarization operators,  $W_A$ and respectively $W_A^{\flat}=W_AN_A$,  defined by  the spectral representations 
\begin{eqnarray}
	W_A&=&\sum_{\sigma}\sigma\left(\Pi_{A\,\sigma}^{(+)} -\Pi^{(-)}_{A\,\sigma} 
	\right)\,,\\	
	W_A^{\flat}&=&\sum_{\sigma}\sigma\left(\Pi_{A\,\sigma}^{(+)} +\Pi^{(-)}_{A\,\sigma} 
	\right)\,,	 
\end{eqnarray}
have the Fourier transforms,  
\begin{eqnarray}
	\hat W_A({\bf p})&=&\sum_{\sigma}\sigma\left(\hat\Pi_{A\,\sigma}^{(+)}({\bf p}) -\hat\Pi^{(-)}_{A\,\sigma}({\bf p})  \right)\nonumber\\
	&=&\sum_{\sigma}2\sigma \left[\frac{m\gamma^0}{E_{p\sigma} }\left( \frac{1}{4}+\sigma\frac{{\bf s}\cdot{\bf p}}{p}\right)+\left(1-2\sigma \frac{\mu_A}{p}\right)\frac{\gamma^0 \boldsymbol \gamma\cdot{\bf p}+2\sigma p \gamma^5 }{4 E_{p\sigma} } \right]\,,\label{q}\\
	\hat W^{\flat}_A({\bf p})&=& \sum_{\sigma}\sigma\left(\hat\Pi_{A\,\sigma}^{(+)}({\bf p}) +\hat\Pi^{(-)}_{A\,\sigma}({\bf p})  \right)=\frac{{\bf s}\cdot{\bf p}}{p}=\hat h({\bf p})\,.
\end{eqnarray}
We observe that the odd polarization operator, $W^{\flat}_A=h$,  is not affected by the presence of chemical potentials while the even one, $W_A$,  depends on $\mu_A$. With these commuting operators we define the operators of particle and antiparticle polarization,
\begin{equation}
	W_A^{pa}=\frac{1}{2}\left( W_A+h   \right)\,,\quad W_A^{ap}=\frac{1}{2}\left( W_A-h   \right)\,,
\end{equation} 
which  are conserved and translation invariant commuting with $H_A$ and $P^i$. They act on the mode spinors (\ref{coco}) and (\ref{moco}) as 
\begin{eqnarray}
	&{W}_A^{pa} U_{A\,{\bf p},\sigma}=\sigma \, U_{A\,{\bf p},\sigma}\,,~~~~~~\quad 
	&	{W}_A^{pa} V_{A\,{\bf p},\sigma}=0\,,\\
	&{W}_A^{ap} U_{A\,{\bf p},\sigma}=0\,,\hspace*{14mm} \quad 
	&	{W}_A^{ap} V_{A\,{\bf p},\sigma}=-\sigma \, V_{A\,{\bf p},\sigma}\,,	
\end{eqnarray} 
pointing out independently the particle and antiparticle polarizations.  In ultra-relativistic regime, when the momentum is extremely  large, we have $p>\mu_A$ and we can take $m=0$. In this limit we find again  
\begin{equation}
	\lim_{m\to 0}	W_A^{pa/ap}=\frac{1}{4}\,\gamma^5\pm \frac{1}{2}\,h\,, 
\end{equation}
as in the free case.

As a general conclusion we  may say that the theory at the level of RQM is difficult involving complicated Fourier transforms as, for example,  in Eqs. (\ref{p}) and (\ref{q}).   Bearing in mind that similar difficulties of  Pryce's theory of spin operator were solved at the level of QFT \cite{Cot}, we shall use all the above results for performing the quantization.

\section{Field and one-particle operators of QFT} 

Here we discuss two systems,  the Dirac free field with $\mu_A=0$ having the mode spinors of the basis ${\frak B}$ and   the Dirac fields with axial chemical potential $\mu_A\not=0$ whose mode spinors form the basis ${\frak B}_A$. The first system   was revisited recently \cite{Cot,Cot1} applying 
the  Bogolyubov method of quantization \cite{Bog} which consists in replacing the wave spinors with field operators  $(\alpha, \alpha^*)\to ({\frak a},{\frak a}^{\dag})$ and $(\beta,\beta^*)\to ({\frak b},{\frak b}^{\dag})$ in the expansion (\ref{Psi}) for obtaining the field operator
\begin{eqnarray}\label{Psix}
	\Psi  (x)=\int d^3p \sum_{\sigma}\left[U_{{\bf p},\sigma}(x) {\frak a}_{\sigma}({\bf p}) +V_{{\bf p},\sigma}(x) {\frak b}^{\dag} _{ \sigma}({\bf p})\right]\,.
\end{eqnarray}
The field operators ${\frak a}$ and  ${\frak b}$ satisfy canonical anti-commutation rules \cite{Cot,Cot1} and define the vacuum state $|0\rangle$ as ${\frak a}|0\rangle={\frak b}|0\rangle =0$.  Moreover, through quantization the  expectation value of any  conserved operator $X$  of RQM becomes  the one-particle  operator, 
$X~\Rightarrow ~ \mathsf{X}=:\langle\Psi  , X\Psi \rangle_D :$ 
calculated  respecting the normal ordering of the operator products \cite{BDR} that changes the sign of the antiparticle term. In this manner we obtained a large algebra of one-particle operators among them we mention the charge operator $\mathsf{Q}=\mathsf{N}^{pa}-\mathsf{N}^{ap}$ and that of the total  number  of particles $\mathsf{N}=\mathsf{N}^{pa}+\mathsf{N}^{ap}$, formed by the particle and antiparticle number operators
\begin{eqnarray}
	\mathsf{N}^{pa}&=&:\langle\psi,\Pi^{(+)} \psi\rangle_D:=\int d^3p\,\sum_{\sigma}{\frak a}_{\sigma}^{\dag}({\vec p}){\frak a}_{\sigma}({\vec p}) \,,	\label{Nplus}\\
	\mathsf{N}^{ap}&=&:-\langle\psi, \Pi^{(-)} \psi\rangle_D:=\int d^3p\, \sum_{\sigma}{\frak b}_{\sigma}^{\dag}({\vec p}){\frak b}_{\sigma}({\vec p})\,.\label{Nminus}
\end{eqnarray} 
The set of  energy and momentum operators,  
\begin{eqnarray}
	\mathsf{H}&=&:\langle\psi, H_D\psi\rangle_D:\nonumber\\
	&=&\int d^3p\,E(p)\sum_{\sigma}\left[{\frak a}_{\sigma}^{\dag}({\vec p}){\frak a}_{\sigma}({\vec p}) +{\frak  b}_{\sigma}^{\dag}({\vec p}){\frak b}_{\sigma}({\vec p})\right]\,,\label{Hom}\\ 
	\mathsf{P}^i&=&:\langle\psi, P^i\psi\rangle_D:\nonumber\\
	&=&\int d^3p\,p^i \sum_{\sigma}\left[{\frak a}_{\sigma}^{\dag}({\vec p}){\frak a}_{\sigma}({\vec p}) +{\frak b}_{\sigma}^{\dag}({\vec p}){\frak b}_{\sigma}({\vec p})\right]\,. \label{Pom}	
\end{eqnarray}
must be completed now with the new particle and antiparticle spin and polarization operators we defined here, 
\begin{eqnarray}
	\mathsf{S}_{i}^{pa}&=&:\langle\Psi  ,S_{i}^{pa}\Psi \rangle_D :\nonumber\\
	&&~~~~~~=\frac{1}{2}\int d^3p\sum_{\sigma,\sigma'}{\Sigma}_{i\,\sigma,\sigma'}({\bf p}) {\frak  a}_{\sigma}^{\dag}({\bf p}){\frak  a}_{\sigma}({\bf p})\,,\\
	\mathsf{S}_{i}^{ap}&=&:\langle\Psi  ,S^{ap}_{i}\Psi \rangle_D :\nonumber\\
	&&~~~~~~=\frac{1}{2}\int d^3p\sum_{\sigma,\sigma'}{\Sigma}_{i\,\sigma,\sigma'}({\bf p}) {\frak  b}_{\sigma}^{\dag}({\bf p}){\frak b}_{\sigma}({\bf p}) \,.\\
	\mathsf{W}^{pa}&=&:\langle\Psi  ,W^{pa}\Psi \rangle_D :
	=\int d^3p\sum_{\sigma}{\sigma} {\frak  a}_{\sigma}^{\dag}({\bf p}){\frak  a}_{\sigma}({\bf p}) \,,\\
	\mathsf{W}^{ap}&=&:\langle\Psi  ,W^{ap}\Psi_A \rangle_D :
	=\int d^3p\sum_{\sigma}{\sigma} {\frak  b}_{\sigma}^{\dag}({\bf p}){\frak  b}_{\sigma}({\bf p}) \,,
\end{eqnarray}
where the matrices $\Sigma_i({\bf p})$  defined by Eq. (\ref{Sig}) are given in the Appendix B.

For the system with $\mu_a\not=0$, when we have the set of the new mode spinors ${\frak B}_A$, we may apply the same method of quantization but considering new field operators $\hat{a}$ and $\hat{\frak b}$ and a new corresponding vacuum state $|0A\rangle$ satisfying $\hat{\frak a}|0A\rangle=\hat{\frak b}|0A\rangle =0$. Then the Dirac  field (\ref{PsiA}) becomes the new field operator  
\begin{eqnarray}\label{Psiq}
	\Psi_A (x)=\int d^3p \sum_{\sigma}\left[U_{A\,{\bf p},\sigma}(x) \hat{\frak a}_{\sigma}({\bf p}) +V_{A\,{\bf p},\sigma}(x) \hat{\frak b}^ {\dag} _{ \sigma}({\bf p})\right]\,,
\end{eqnarray}  
acting on  the same Fock state as $\Psi$. In this case the conserved one-particle operators may be derived  using obvious identities as 
\begin{eqnarray}
	:\langle\Psi_A  , \Pi^{(+)}_{A\,\sigma}\Psi_A \rangle_D : &=&\int d^3p\, \hat{\frak  a}_{\sigma}^{\dag}({\bf p})\hat{\frak  a}_{\sigma}({\bf p})	\,,\\
	:\langle\Psi_A  , \Pi^{(-)}_{A\,\sigma}\Psi_A \rangle_D : &=&-\int d^3p\, \hat{\frak  b}_{\sigma}^{\dag}({\bf p})\hat{\frak  b}_{\sigma}({\bf p})	\,,
\end{eqnarray}
for deriving the charge operator 
\begin{eqnarray}
	\mathsf{Q}_A&=&:\langle\Psi_A  ,\Psi_A \rangle_D :\nonumber \\
	& =&\int d^3p\sum_{\sigma}\left[ \hat{\frak  a}_{\sigma}^{\dag}({\bf p})\hat{\frak  a}_{\sigma}({\bf p}) -\hat{\frak  b}_{\sigma}^{\dag}({\bf p})\hat{\frak  b}_{\sigma}({\bf p})  \right]\,,
\end{eqnarray}
the operator of number of particles,
\begin{eqnarray}
	\mathsf{N}_A&=&:\langle\Psi_A  ,N_A\Psi_A \rangle_D :\nonumber \\
	& =&\int d^3p\sum_{\sigma}\left[ \hat{\frak  a}_{\sigma}^{\dag}({\bf p})\hat{\frak  a}_{\sigma}({\bf p})+\hat{\frak  b}_{\sigma}^{\dag}({\bf p})\hat{\frak  b}_{\sigma}({\bf p})  \right]\,,
\end{eqnarray}
as well as the Hamiltonian and momentum operators
\begin{eqnarray}
	\mathsf{H}_A&=&:\langle\Psi_A  ,H_A\Psi_A \rangle_D :\nonumber \\
	& =&\int d^3p\sum_{\sigma} E_{p\sigma} \left[ \hat{\frak  a}_{\sigma}^{\dag}({\bf p})\hat{\frak  a}_{\sigma}({\bf p}) +\hat{\frak  b}_{\sigma}^{\dag}({\bf p})\hat{\frak  b}_{\sigma}({\bf p})  \right]\,,\\	
	\mathsf{P}_A^i&=&:\langle\Psi_A  ,-i\partial_i \Psi_A \rangle_D :\nonumber \\
	& =&\int d^3p p^i \sum_{\sigma} \left[ \hat{\frak  a}_{\sigma}^{\dag}({\bf p})\hat{\frak  a}_{\sigma}({\bf p}) +\hat{\frak  b}_{\sigma}^{\dag}({\bf p})\hat{\frak  b}_{\sigma}({\bf p})  \right]\,,	
\end{eqnarray}
The  particle and antiparticle spin and  polarization operators we derive here for the first time lead to the one particle operators,
\begin{eqnarray}
	\mathsf{S}_{A\,i}^{pa}&=&:\langle\Psi_A  ,S_{A\,i}^{pa}\Psi_A \rangle_D :\nonumber\\
	&&~~~~~~=\frac{1}{2}\int d^3p\sum_{\sigma,\sigma'}{\Sigma}_{i\,\sigma,\sigma'}({\bf p}) \hat{\frak  a}_{\sigma}^{\dag}({\bf p})\hat{\frak  a}_{\sigma}({\bf p})\,,\\
	\mathsf{S}_{A\,i}^{ap}&=&:\langle\Psi_A  ,S^{ap}_{A\,i}\Psi_A \rangle_D :\nonumber\\
	&&~~~~~~=\frac{1}{2}\int d^3p\sum_{\sigma,\sigma'}{\Sigma}_{i\,\sigma,\sigma'}({\bf p}) \hat{\frak  b}_{\sigma}^{\dag}({\bf p})\hat{\frak b}_{\sigma}({\bf p}) \,.\\
	\mathsf{W}_A^{pa}&=&:\langle\Psi_A  ,W^{pa}_A\Psi_A \rangle_D :
	=\int d^3p\sum_{\sigma}{\sigma} \hat{\frak  a}_{\sigma}^{\dag}({\bf p})\hat{\frak  a}_{\sigma}({\bf p}) \,,\\
	\mathsf{W}_A^{ap}&=&:\langle\Psi_A  ,W^{ap}_A\Psi_A \rangle_D :
	=\int d^3p\sum_{\sigma}{\sigma} \hat{\frak  b}_{\sigma}^{\dag}({\bf p})\hat{\frak  b}_{\sigma}({\bf p}) \,.
\end{eqnarray}
which have very similar expressions as those of the free theory. However, the crucial difference is that now the field operators $\hat{\frak a}$ and $\hat{ \frak b}$ are different from those of the free field, ${\frak a}$ and ${ \frak b}$.

We can verify that this is true assuming that at the instant $t_0$ the field $\psi_A(x)$ satisfies the initial condition 
\begin{equation}\label{ini}
	\left.\Psi_A(t,{\bf x})\right|_{t=t_0}=	\left.\Psi_h(t,{\bf x})\right|_{t=t_0}\,,
\end{equation}
where $\Psi_h$ is the quantum Dirac free field expanded in the basis  ${\frak B}_h$. Using the inversion formulas (\ref{inv}) we may write the field operators of the Dirac field with axial chemical potential as
\begin{eqnarray}
	\hat{\frak a}_{\sigma}({\bf p})	&=&\langle	U_{A\,{\bf p},\sigma}, \Psi_h\rangle_D|_{t=t_0}\nonumber\\
	&=&\int d^3p'\sum_{\sigma'}\left[\langle U_{A\,{\bf p},\sigma}, U_{h\,{\bf p}',\sigma'} \rangle_D|_{t=t_0}  {\frak a}_{\sigma'} ({\bf p}')\right.  \nonumber\\
	&&\left.\hspace*{14mm}+\langle U_{A\,{\bf p},\sigma}, V_{h\,{\bf p}',\sigma'} \rangle_D|_{t=t_0}  {\frak b}^{\dag}_{\sigma'} ({\bf p}')\right]\,,\label{B1}\\
	\hat{\frak b}_{\sigma}({\bf p})&=&\langle  \Psi_h, V_{A\,{\bf p},\sigma}\rangle_D|_{t=t_0}\nonumber\\
	&=&\int d^3p'\sum_{\sigma'}\left[\langle U_{h\,{\bf p}',\sigma'}, V_{A\,{\bf p},\sigma} \rangle_D|_{t=t_0}  {\frak a}^{\dag}_{\sigma'} ({\bf p}')\right.  \nonumber\\
	&&\left.\hspace*{14mm}+\langle V_{h\,{\bf p}',\sigma'}, V_{A\,{\bf p},\sigma} \rangle_D|_{t=t_0}  {\frak b}_{\sigma'} ({\bf p}')\right]\,,\label{B2}
\end{eqnarray}
obtaining a non-trivial Bogolyubov transformation 
\begin{eqnarray}
	\hat{\frak a}_{\sigma}({\bf p}) &=&  A({\bf p},\sigma)  {\frak a}_{\sigma}({\bf p}) + B({\bf p}, \sigma) {\frak b}^{\dag}_{\sigma}(-{\bf p})  \,,\label{Bo1}\\
	\hat{\frak b}_{\sigma}({\bf p}) &=& A({\bf p},\sigma)  {\frak b}_{\sigma}({\bf p}) + B({\bf p}, \sigma) {\frak a}^{\dag}_{\sigma}(-{\bf p})  \,,\label{Bo2}
\end{eqnarray}
whose coefficients,  derived in the Appendix C, satisfy the conditions $A(-{\bf p},\sigma)=A({\bf p},\sigma)$, $B(-{\bf p},\sigma)=-B({\bf p},\sigma)$ and
\begin{equation}
	A({\bf p},\sigma)^2+|B({\bf p},\sigma)|^2=1\,,	
\end{equation} 
which guarantees that the field operators satisfy canonical anticommutation rules. We remind the reader that $|B({\bf p},\sigma)|^2$ gives the probability of measuring free particles in the vacuum state of the system with axial chemical potential, $|0A\rangle$.

On the other hand, the initial condition (\ref{ini}) allows us to relate the field operators $\Psi_h$ and $\Psi_A$ at any time $t$, taking into account that these fields evolve as
\begin{eqnarray}
	\Psi_h(t,{\bf x})&=&e^{i\mathsf{H}(t-t_0)}	\Psi_h(t_0,{\bf x})e^{-i\mathsf{H}(t-t_0)}\,,\\
	\Psi_A(t,{\bf x})&=&e^{i\mathsf{H}_A(t-t_0)}	\Psi_A(t_0,{\bf x})e^{-i\mathsf{H}_A(t-t_0)}\,.		
\end{eqnarray}	
Hereby we obtain  the unitary transformation
\begin{equation}
	\Psi_A(t,{\bf x})=\mathsf{U}(t-t_0)\Psi_h(t,{\bf x})\mathsf{U}^{\dag}(t-t_0)\,,	
\end{equation}
whose  unitary operator 
$\mathsf{U}(t)=	e^{i\mathsf{H}_At} e^{-i\mathsf{H}t}$ 
assures the equivalence between the time evolution pictures governed by the Hamiltonian operators $\mathsf{H}$ and  the  $\mathsf{H}_A$.

\section{Concluding remarks}

We studied here systematically the Dirac theory with chemical potentials focusing on the new conserved operators that can be constructed using spectral representations. For applying this method we solved first the Dirac equation with axial chemical potential obtaining new mode spinors and corresponding projection operators. As in the case of the free field, we focused on the spin and polarization operators showing how  the helicity and our even polarization operator \cite{Cot} may be combined for obtaining the particle and antiparticle polarization operators. An intermediate step of this construction is the definition of the odd spin-type operator whose components enlarge the $su(2)$ spin algebra to a $so(4)$ one that can be split in two $su(2)$ particle and antiparticle spin algebras that now can be considered separately.  

After quantization,  we obtained the one-particle operators of  Dirac's QFT with mass and axial chemical potential we need for writing down the statistical operator  
\begin{equation}
	\rho_A=\frac{1}{Z}\exp\beta(-\mathsf{H}_A+\mu\mathsf{N}_A+\mu_V\mathsf{Q}_A+\mu_{pa}\mathsf{W}_A^{pa} +\mu_{ap}\mathsf{W}_A^{ap})\,,
\end{equation}
in which we may introduce different particle and antiparticle helical chemical potentials, $\mu_{pa}$ and $\mu_{ap}$.

\section*{Appendx A: Dirac representation }

\renewcommand{\theequation}{A\arabic{equation}}
\setcounter{equation}{0} \renewcommand{\theequation}
{A.\arabic{equation}}

We use here the Dirac $\gamma$-matrices in the standard representation with diagonal $\gamma^0$. These satisfy $\{\gamma^{\mu},\gamma^{\nu}\}=2\eta^{\mu\nu}$  giving rise to the  generators
$s^{\mu\nu}=\frac{i}{4}\left[\gamma^{\mu},\gamma^{\nu}\right]=\overline{s^{\mu\nu}}$ of the Dirac reducible representation $\rho_D=(\frac{1}{2},0)\oplus(0,\frac{1}{2})$ of the $SL(2,\mathbb{C})$ group in the four-dimensional space ${\cal V}_D={\cal V}_P\oplus{\cal V}_P$ of  Dirac spinors.  All these matrices, including the $SL(2,\mathbb{C})$ generators, are Dirac self-adjoint such that the transformations
\begin{equation}\label{tr}
	\lambda(\omega)=\exp\left(-\frac{i}{2}\omega^{\alpha\beta}s_{\alpha\beta}\right)\in \rho_D[SL(2,\mathbb{C})]\,, 
\end{equation}
having  real-valued parameters, $\omega^{\alpha \beta}=-\omega^{\beta\alpha}$,   leave invariant the Hermitian form $\overline{\Psi_A }\Psi_A $  as $\overline{\lambda(\omega)}=\lambda^{-1}(\omega)=\lambda(-\omega)$. 

We denote  by 
\begin{equation}\label{r0}
	r={\rm diag}(\hat r,\hat r)\in {\rho_D}\left[SU(2)\right]	
\end{equation}
the transformations we call here simply rotations  for which we use the  generators  
\begin{equation}\label{si}
	s_i= \frac{1}{2}\epsilon_{ijk}s^{jk}  ={\rm diag}(\hat s_i,\hat s_i)=-\frac{1}{2}\gamma^0\gamma^5\gamma^i\,, \quad \hat s_i=\frac{1}{2}\sigma_i\,,
\end{equation}
expressed in terms of Pauli matrices $\sigma_i$, and Cayley-Klein parameters  $\theta^i=\frac{1}{2}\epsilon_{ijk}\omega^{jk}$ such that
\begin{eqnarray}
	r(\theta)={\rm diag}(\hat r(\theta),\hat r(\theta))\,,\quad~~ \hat r(\theta)=e^{-i \theta^i \hat s_i}=e^{-\frac{i}{2} \theta^i \sigma_i}\,. \label{r}
\end{eqnarray}
Similarly, we chose the parameters  $\tau^i=\omega^{0i}$ and the generators 
\begin{equation}\label{s0i}
	s_{i0}=s^{0i} =\frac{i}{2}\gamma^0\gamma^i\,,
\end{equation}
for the Lorentz boosts,  
\begin{eqnarray}
	l(\tau)=e^{-i \tau^i s_{0i}} \,.\label{l}
\end{eqnarray}
Of a special interest are the Wigner  boosts  with parameters $\tau^i=-\frac{p^i}{p}{\rm tanh}^{-1} \frac{p}{E_p}$   \cite{Th},
\begin{eqnarray}\label{Ap}
	l_{{\bf p}}=\frac{E_p+m+\gamma^0{\boldsymbol\gamma}\cdot {\bf p}}{\sqrt{2m(E_p+m)}}=	l_{{\bf p}}^+\,,\quad l^{-1}_{\bf p}=l_{-{\bf p}}=\bar{l}_{\bf p}\,,
\end{eqnarray}
which enter in the structure of the Dirac mode spinors.

\section*{ Appendix B: Polarization spinors}

\setcounter{equation}{0} \renewcommand{\theequation}
{B.\arabic{equation}}

The  polarization Pauli spinors, $\xi_{\sigma}({\bf n})$,  and   $\eta_{\sigma}({\bf n})=i\sigma_2 \xi_{\sigma}^*({\bf n})$ which satisfy the eigenvalue problems (\ref{xy})  have the form  
\begin{eqnarray}
	\xi_{\frac{1}{2}}({\bf n})&=&\sqrt{\frac{1+n^3}{2}}\left(
	\begin{array}{c}
		1\\
		\frac{n^1+i n^2}{1+n^3}
	\end{array}\right)\,, \nonumber\\
	\xi_{-\frac{1}{2}}({\bf n})&=&\sqrt{\frac{1+n^3}{2}}\left(
	\begin{array}{c}
		\frac{-n^1+i n^2}{1+n^3}\\
		1
	\end{array}\right)\,,\label{xi1}
\end{eqnarray}
representing  related  orthonormal,   
\begin{equation}\label{xyort}
	\xi_{\sigma}^+({\bf n})\xi_{\sigma'}({\bf n})=\eta_{\sigma}^+({\bf n})\eta_{\sigma'}({\bf n})=\delta_{\sigma\sigma'}\,,	
\end{equation}
and complete systems, 
\begin{equation}\label{xycom}
	\sum_{\sigma}\xi_{\sigma}({\bf n})\xi_{\sigma}^+({\bf n})=\sum_{\sigma}\eta_{\sigma}({\bf n})\eta_{\sigma}^+({\bf n})=1_{2\times2}\,,	
\end{equation}
which form bases in the subspaces of Pauli spinors,  ${\cal V}_P$, of the space of Dirac spinors, ${\cal V}_D$. 

For writing down the spin components (\ref{SSS}) in this basis we  use the matrices  \cite{Cot}, 
\begin{eqnarray} 
	\Sigma_{1}({\bf n})&=&{n^1}\,\sigma_3 -n^1\frac{n^1\sigma_1+n^2\sigma_2}{1+n^3}+\sigma_1\,,\nonumber\\
	\Sigma_{2}({\bf n})&=&{n^2}\,\sigma_3 -n^2\frac{n^1 \sigma_1+n^2\sigma_2}{1+n^3}+\sigma_2\,,\nonumber \\
	\Sigma_{3}({\bf n})	&=&{p^3}\sigma_3-(p^1 \sigma_1+p^2\sigma_2) \,,\label{Sigp1}
\end{eqnarray}
which satisfy the condition ${n}^i{\Sigma}_i({\bf n})=\sigma_3$.

\section*{Appendix C: Bogolyubov coefficients}

\setcounter{equation}{0} \renewcommand{\theequation}
{C.\arabic{equation}}

Eqs. (\ref{B1}) and (\ref{B2}) define a Bogolyubov transformation  of general form
\begin{eqnarray}
	\hat{\frak a}_{\sigma}({\bf p}) &=&\sum_{\sigma'} \left[ A_{\sigma,\sigma'}({\bf p}) {\frak a}_{\sigma'}({\bf p}) + B_{\sigma,\sigma'}({\bf p}) {\frak b}^{\dag}_{\sigma'}(-{\bf p})    \right]\,,\label{Bo1}\\
	\hat{\frak b}_{\sigma}({\bf p}) &=&\sum_{\sigma'} \left[ D_{\sigma,\sigma'}({\bf p}) {\frak b}_{\sigma'}({\bf p}) + C_{\sigma,\sigma'}({\bf p}) {\frak a}^{\dag}_{\sigma'}(-{\bf p})    \right]\,,\label{Bo2}
\end{eqnarray} 
whose coefficients can be identified as
\begin{eqnarray}
	&&\langle U_{A\,{\bf p},\sigma}, U_{h\,{\bf p}',\sigma'}\rangle_D|_{t=t_0}= \delta^3({\bf p}-{\bf p}') A_{\sigma,\sigma'}({\bf p})\nonumber\\
	&&~~~~~~=\delta^3({\bf p}-{\bf p}')e^{i t_0(E_{p\sigma}-E_p)}u_A^+({\bf p},\sigma)u({\bf p},\sigma')\,,\label{in}\\
	&&\langle U_{A\,{\bf p},\sigma}, V_{h\,{\bf p}',\sigma'} \rangle_D|_{t=t_0}= \delta^3({\bf p}+{\bf p}') B_{\sigma,\sigma'}({\bf p}) \nonumber\\
	&&~~~~~~=\delta^3({\bf p}+{\bf p}')e^{i t_0(E_{p\sigma}+E_p)}u_A^+({\bf p},\sigma)v(-{\bf p},\sigma')\,,\\
	&&\langle U_{h\,{\bf p}',\sigma'}, V_{A\,{\bf p},\sigma} \rangle_D|_{t=t_0}= \delta^3({\bf p}+{\bf p}') C_{\sigma,\sigma'}({\bf p})\nonumber\\
	&&~~~~~~=\delta^3({\bf p}+{\bf p}')e^{i t_0(E_{p}+E_{p\sigma})}u^+(-{\bf p},\sigma')v_A({\bf p},\sigma)\,,\\
	&&\langle V_{h\,{\bf p}',\sigma'}, V_{A\,{\bf p},\sigma} \rangle_D|_{t=t_0}= \delta^3({\bf p}-{\bf p}') D_{\sigma,\sigma'}({\bf p})\nonumber	\\
	&&~~~~~~=\delta^3({\bf p}-{\bf p}')e^{i t_0(E_p-E_{p\sigma})}v^+({\bf p},\sigma')v_A({\bf p},\sigma)\,,\label{out}
\end{eqnarray}  
Setting for simplicity  $t_0=0$ we obtain the expressions
\begin{eqnarray}
	A_{\sigma,\sigma'}({\bf p})&=&u_A^+({\bf p},\sigma)u_h({\bf p},\sigma')=\delta_{\sigma\sigma'}   { A}({\bf p},\sigma)\,,\nonumber \\
	B_{\sigma,\sigma'}({\bf p})&=&u_A^+({\bf p},\sigma)v_h(-{\bf p},\sigma')=\delta_{\sigma\sigma'}   {B}({\bf p},\sigma)\,, \nonumber \\
	C_{\sigma,\sigma'}({\bf p})&=& u_h^+(-{\bf p},\sigma')v_A({\bf p},\sigma) =\delta_{\sigma\sigma'}   {B}({\bf p},\sigma)\,, \nonumber \\
	D_{\sigma,\sigma'}({\bf p})&=& v_h^+({\bf p},\sigma')v_A({\bf p},\sigma)=\delta_{\sigma\sigma'}    { A}({\bf p},\sigma)\,,\nonumber
\end{eqnarray} 
in which we substitute the spinors of the momentum-helicity basis, (\ref{uh}) and (\ref{vh}), as well as those for $\mu_A\not=0$,  (\ref{coco}) and (\ref{moco}),  obtaining
\begin{eqnarray}
	{A}({\bf p},\sigma)&=&\frac{(E_p+m)(E_{p\,\sigma}+E_p)-2\sigma p\mu_A}{2\sqrt{E_pE_{p\,\sigma}(E_p+m)(E_{p\,\sigma}+m)}}\,,\\
	{B}({\bf p},\sigma)&=&-\frac{2\sigma p(E_{p\,\sigma}-E_p)+\mu_A(E_p+m)}{2\sqrt{E_pE_{p\,\sigma}(E_p+m)(E_{p\,\sigma}+m)}}e^{-2i\sigma \phi({\bf p})},~~~~	\label{Bps}
\end{eqnarray}		
where 	$E_p=\sqrt{p^2+m^2}$,  $E_{p\,\sigma}$ is given by Eq. (\ref{Eps}) while $\phi({\bf p})	=\arctan\frac{p^2}{p^1}$. Note that we  used here the orthonormalization relations (\ref{xyort}) written for ${\bf n}={\bf n}_p=\frac{\bf p}{p}$ and the identities
\begin{eqnarray}
	\xi^+_{\sigma}({\bf p})\eta_{\sigma'}(-{\bf p})=-\delta_{\sigma\sigma^{\prime}}\frac{p^1-2i\sigma p^2}{\sqrt{{p^1}^2+{p^2}^2}}=-\delta_{\sigma\sigma^{\prime}} e^{-2i\sigma \phi({\bf p})}\,.\nonumber
\end{eqnarray}
Hereby we see that the Bogolyubov transformation under consideration  is non-trivial for $\mu_A\not=0$.

\end{document}